\documentclass[conference]{IEEEtran}

\usepackage{array}
\usepackage{subfigure}
\usepackage{algorithm}
\usepackage{algpseudocode}
\usepackage{graphicx}
\usepackage{amssymb}
\usepackage{amsmath}
\usepackage{cite}
\usepackage{amsmath}
\usepackage{float}
\usepackage[switch]{lineno}
\usepackage{color}

\ifCLASSINFOpdf
\else
\fi

\begin{document}

\renewcommand{\IEEEQED}{\IEEEQEDopen}
\newtheorem{lemma}{\textbf{Lemma}}
%
\title{Duty-Cycle-Aware Minimum-Energy Multicasting in Wireless Sensor Networks}

\author{\IEEEauthorblockN{Kai Han}
\IEEEauthorblockA{School of Computer Science\\
Zhongyuan University of Technology\\
Zhengzhou, China 450007\\
Email: hankai@gmail.com}
\and
\IEEEauthorblockN{Yang Liu}
\IEEEauthorblockA{School of Information Sciences
and Engineering\\Henan University of Technology\\
Zhengzhou, China 450001\\
Email: enjoyang@gmail.com}
\and
\IEEEauthorblockN{Jun Luo}
\IEEEauthorblockA{School of Computer Engineering\\Nanyang Technological University\\
Singapore\\
Email: junluo@ntu.edu.sg}}

\maketitle

\begin{abstract}
In duty-cycled wireless sensor networks, the nodes switch between active and
dormant states, and each node may determine its active/dormant schedule
independently. This complicates the Minimum-Energy Multicasting (MEM)
problem, which has been primarily studied in always-active wireless ad-hoc
networks. In this paper, we study the duty-cycle-aware MEM problem in
wireless sensor networks, and we present
a formulation of the Minimum-Energy Multicast Tree Construction and
Scheduling (MEMTCS) problem. We prove that the MEMTCS problem is
NP-hard, and it is unlikely to have an approximation algorithm with a
performance ratio of $(1-o(1))\ln\Delta$, where $\Delta$ is the maximum node
degree in a network. We propose a polynomial-time approximation algorithm
for the MEMTCS problem with a performance ratio of $\mathcal{O}(H(\Delta+1))$, where $H(\cdot)$ is the harmonic number. We also provide a distributed implementation of our algorithm. Finally, we perform extensive simulations and the results demonstrate that our algorithm significantly outperform other known algorithms in terms of both the total energy cost and the transmission redundancy.
\end{abstract}

\IEEEpeerreviewmaketitle

\section{Introduction}
Wireless Sensor Networks (WSNs) are decentralized wireless
networks without any preexisting infrastructures, and the sensor nodes are usually powered by batteries. Since the limited battery lifetime imposes a severe constraint on
the network performance, it is imperative to develop energy conservation
mechanisms for WSNs. One common approach for energy conservation in WSNs is
duty-cycling, in which each node switches between active and dormant states,
and the active/dormant schedule can differ from node to node \cite{Gu2007,Anastasi2009,Wang2009,Guo2009,Su2009,Hong2010,Xiong2011}.
Duty-cycling can be easily implemented, and is proved to be an
effective way for energy conservation\cite{Anastasi2009}. Consequently, Duty-Cycled Wireless Sensor Networks (DC-WSNs) have often been adopted by various applications \cite{Gui2004,He2006,Mo2009}.

Multicasting is a crucial component of wireless networking, and it has been applied to WSNs in supporting data dissemination for distributed data management (e.g., \cite{Xing2009}). Therefore, designing an energy efficient multicast protocol is of great importance. In
an Always-Active Wireless Ad-hoc NETwork (AA-WANET), the network topology is static, and each forwarding node can cover all its neighboring nodes by only one transmission. Therefore, the main task of the Minimum-Energy Multicasting (MEM) problem in AA-WANETs is to select appropriate forwarding nodes such that a multicast tree with the minimum energy cost can be constructed. This problem was proved to be NP-hard, and some approximation algorithms have been proposed \cite{Wieselthier2000,Wan2004,Liang2006,Li2007,Liang2009}.

In DC-WSNs, however, new challenges arise. More specifically, the network topology is now intermittently connected, and a forwarding node may transmit the same data
packet many times to reach its neighboring nodes. Therefore, to design
energy efficient multicasting algorithms in DC-WSNs, not only the
forwarding nodes should be selected appropriately to construct a multicast tree, but also the transmissions of each forwarding node need to be scheduled
intelligently to cover the receiver nodes and reduce the transmission redundancy. These two
problems must be handled holistically so that the total energy cost
can be reduced. Thus, the existing solutions for the MEM problem
in AA-WANETs are not suitable for DC-WSNs, and we need to design new energy
efficient multicasting algorithms to meet the challenges in DC-WSNs.

\subsection{Background and Motivations}
The MEM problem in wireless ad-hoc networks has been studied in \cite{Wieselthier2000,Wan2004,Liang2006,Li2007,Liang2009}.
Wieselthier \textit{et al.}\cite{Wieselthier2000} studied the minimum power broadcast/multicast routing problems under a scenario
where each node can adjust its transmission power continuously, and proposed several greedy heuristics.
Wan \textit{et al.}\cite{Wan2004} proved that the heuristics proposed by \cite{Wieselthier2000} have linear approximation
ratios, and provided several approximation algorithms with constant approximation ratios for the MEM problem based on the Minimum Steiner Tree algorithm.
Liang \cite{Liang2006} considered a scenario in which each wireless node can adjust its
transmission power in a discrete fashion, and the communication links are
symmetric. He proposed a centralized approximation algorithm with performance
ratio $4\ln|M|$ for building a minimum-energy multicasting tree, where $M$
is the set of destination nodes in a multicast request. Li \textit{et al.}\cite{Li2007}
considered a case in which all nodes have a fixed transmission power and the
communication links are asymmetric. They converted the minimum-energy
multicasting problem to an instance of the Directed Steiner Tree (DST)
problem\cite{Charikar1998}, and presented several heuristics.
Liang \textit{et al.}\cite{Liang2009} studied the minimum-energy all-to-all multicasting problem in wireless ad-hoc networks, which aims at building a shared multicast tree such that the total energy consumption of performing an all-to-all multicast session by this tree is minimized. They proved the NP-completeness of this problem and also proposed several approximation algorithms. However, all the algorithms proposed in \cite{Li2007,Liang2006,Wieselthier2000,Liang2009,Wan2004} took \textbf{the assumption that the
network nodes are always-active}. Therefore, they are not directly applicable to DC-WSNs.

Recently, the \textbf{data dissemination problems in DC-WSNs} have started to attract attentions from the research community. Wang and Liu \cite{Wang2009} tackled the broadcast scheduling problem in DC-WSNs based on the dynamic-programming approach. Guo \textit{et al}.\ \cite{Guo2009} considered the effect of unreliable links on
broadcasting, and proposed an opportunistic forwarding scheme to reduce the
broadcast delay and redundancy in DC-WSNs. Hong \textit{et al}.\ \cite{Hong2010} studied the
Minimum-Transmission Broadcasting (MTB) problem in DC-WSNs. They adopted a
\textbf{restricted duty-cycling model}, where only one active time slot exists
in a working period of each node. They proposed a centralized algorithm
with approximation ratio of $3(\ln\Delta+1)$ and a distributed algorithm
with approximation ratio of 20. Note that under their particular duty-cycling model, any node can determine its optimal transmission
schedule in polynomial-time for transmitting a data packet to a set of neighboring nodes. Therefore, \textbf{their methods cannot be adapted to our case}. Xiong \textit{et al}.\ \cite{Xiong2011} adopted the same duty-cycling model as that in \cite{Hong2010}, and studied the load balancing problem for data dissemination in DC-WSNs. They proposed several scheduling algorithms for data transmissions such that the maximum workload of the forwarding nodes is minimized. However, they assumed that the routing paths are pre-computed, and their concern is load balancing instead of energy-efficiency.

To the best of our knowledge, the only work which studied the MEM problem in
DC-WSNs is \cite{Su2009}. In \cite{Su2009}, the authors adopted another \textbf{restricted duty-cycling model} in which
the active time slots of any node must be consecutive, and
proposed two optimal algorithms (``oCast" and ``DB-oCast") for the
MEM problem in DC-WSNs. As the
energy cost for receiving data was neglected in \cite{Su2009}, the
algorithms proposed in \cite{Su2009} actually minimize the transmission redundancy in a multicast session. Most importantly, although oCast and DB-oCast were both claimed to be optimal in \cite{Su2009}, \textbf{their time complexity grows exponentially} with respect to the number of destination nodes.

\subsection{Our Contributions}

In this paper, we study the MEM problem in DC-WSNs using a \textbf{generic duty-cycling model}, where each wireless node determines its active/dormant
schedule without any constraints.
We formulate the MEM problem for DC-WSNs and prove its NP-hardness. We
propose an approximation algorithm with guaranteed performance ratio, as well as a distributed implementation of our algorithm.
The contributions of our work can be summarized as follows:
\renewcommand{\labelenumi}{\arabic{enumi})}
\begin{enumerate}
\item We formulate the Minimum-Energy Multicast
Tree Construction and Scheduling (MEMTCS) problem and prove its NP-hardness.
We also prove that, unless $\mathit{NP} \subseteq \mathit{DTIME}(n^{\mathcal{O}(\log\log n)})$, the
MEMTCS problem cannot be approximated with a performance ratio of $%
(1-o(1))\ln\Delta$, where $\Delta$ is the maximum node degree of the network graph under consideration.
\item We propose a polynomial-time approximation algorithm for the MEMTCS
problem with an approximation ratio of $12\rho H(\Delta+1)+4\rho$, where $H(\cdot)$
is the harmonic number and $\rho$ is the approximation ratio of a given
algorithm for the Minimum Steiner Tree (MST) problem.
\item We present a distributed implementation of the proposed algorithm, and we conduct extensive simulations to evaluate the performance of our algorithm. The simulation results demonstrate that our algorithm
significantly outperform other known algorithms in terms of both the total
energy cost and the transmission redundancy.
\end{enumerate}
To the best of our knowledge, we are the first to present a polynomial-time
approximation algorithm with provable approximation ratio for the MEM
problem in DC-WSNs.

The rest of our paper is organized as follows. In Section II, we introduce
the wireless network model and formulate the MEMTCS problem. In Section III, we analyze the complexity of MEMTCS, and
we propose an approximation algorithm for it. A distributed implementation of the proposed algorithm is presented in Section IV. In Section V, we evaluate the performance of the proposed algorithm by simulations. Section VI concludes the paper. In order to maintain fluency, we postpone the proofs to the appendix.

\section{Assumptions and Definitions}
In this section, we first describe our network model and related parameters, then we present the formulation of the MEMTCS problem that we tackle in this paper.

\subsection{Network Model and Parameters}

A wireless sensor network is modeled by an undirected graph $G=(V,E)$, where
$V$ is the set of wireless nodes, and $E$ is the set of links. The nodes in $%
V$ are distributed in a two-dimensional plane and each node is equipped with
an omni-directional antenna. We assume that all nodes have the same fixed
transmission power, and there exists a link between two nodes if they are
within the transmission range of each other. We also assume that each node has an unique ID and knows the IDs of its one-hop neighbors.

We assume that time is divided into equal-length slots, and each time slot
is long enough for sending or receiving a data packet. Without loss of
generality, we assume that the working schedule of each node is periodic,
and the \textit{working period} of any node has $K$ time slots. To save energy, every
node switches between active and dormant states, and we denote by $\Gamma(u)$
the set of active time slots in the working period of node $u$ ($\Gamma
(u)\subseteq \{ 1,2,...,K\}$ and $\Gamma (u) \ne \emptyset$) . The
active/dormant schedule is independently determined by each node without any
constraints. Note that our duty-cycling model is similar with the model used
in \cite{Guo2009,Gu2007}, and the duty-cycling models used in \cite{Hong2010,Su2009,Xiong2011} can be considered as
special cases of our model. Following a very common setting in DC-WSNs, we
assume that \textbf{a node can wake up its transceiver to transmit a packet
at any time slot, but can only receive a packet when it is active}. We also
assume that the time synchronization is achieved in network, and each node knows the active/dormant schedule of its neighboring nodes. These are common assumptions in the literature \cite{Guo2009,Hong2010,Su2009,Gu2007,Wang2009,Xiong2011}.

We denote by ${e_s}$ the energy cost for sending a data packet by any node, and
denote by ${e_r}$ the energy cost for receiving a data packet. It is well known that a wireless sensor node has different power consumption levels at different working states such as transmitting, receiving, and idle-listening. Let $e_{tx}$, $e_{rx}$, and $e_{il}$ be the energy consumptions of radio for transmitting, receiving and idle-listening, respectively. Usually, $e_{tx}$ is larger than $e_{rx}$, whereas $e_{il}$ is only slightly smaller than $e_{rx}$.
Since a node has to consume at least $e_{il}$ when it is active, we set ${e_s}$ to be the value of $e_{tx} - e_{il}$, and set ${e_r}$ to be the value of $e_{rx} - e_{il}$. Clearly, ${e_s}\ge{e_r}\ge 0$.

For the convenience of description, we clarify some other notations here. For any node $u$, we denote by ${nb_G}(u)$ the set of neighboring
nodes of $u$ in $G$. Suppose that $T$ is an arbitrary tree. We denote by $%
N(T)$ the set of nodes in $T$. Denote by $E(T)$ the set of edges in $T$.
Denote by ${d^1}(T)$ the set of nodes in $T$ with degree one. Denote by ${d^+%
}(T)$ the set of nodes in $T$ with degree greater than one. If $T$ is a
rooted tree, then we denote by $nl(T)$ the set of non-leaf nodes in $T$, and
denote by $child(u,T)$ the set of child nodes of node $u$ in $T$. Suppose
that $r$ is the root node of $T$. Clearly, if $r\in {d^1}(T)$, then $nl(T)={%
d^+}(T)\cup\{r\}$; otherwise $nl(T)={d^+}(T)$.

\subsection{The MEMTCS Problem} \label{sec:memtcs}

In a multicast session, there exists a \textit{terminal set} $M\subseteq V$
and a source node $s\in M$, such that the data sent by $s$ should be
received by all the nodes in $M-\{s\}$. A multicast tree $T$ is a sub-tree
of $G$ which is rooted at $s$, and each terminal node in $M$ is a tree node
of $T$.

Since all nodes have the same transmission power, the energy cost for
sending a data packet using a multicast tree $T$ can be determined by the
number of total transmissions of the nodes in $\mathit{nl}(T)$. In DC-WSNs, the time
slot at which a node transmits the packet decides which neighbors can
receive it. Therefore, the transmission schedule of each forwarding node plays
a key role on the energy cost for multicasting. Furthermore, incorrect
schedules can even prevent the destination nodes from receiving a data
packet.

\begin{figure}[htbp]
\centerline{\includegraphics[scale=0.9]{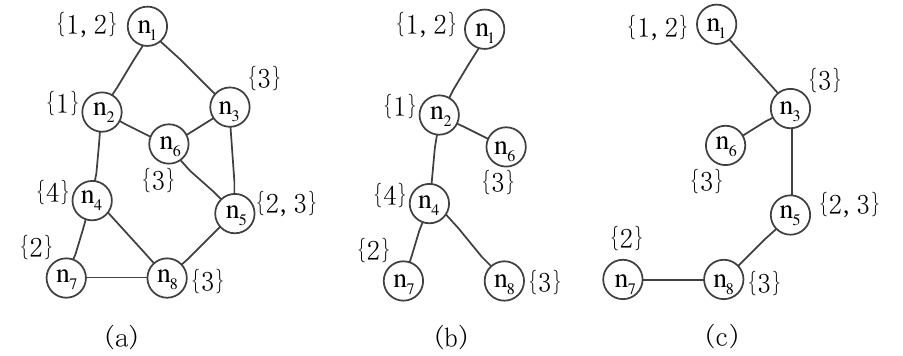}}
\label{DCWSNExample}
\centering
\caption{A DC-WSN graph $G$ (a) and two multicast trees $T_1$(b) and $T_2$(c).}
\label{fig:example}
\end{figure}

For example, consider the wireless network shown in Fig.~\ref{fig:example}(a). Suppose that
the terminal set is $\{n_1,n_6,n_7,n_8\}$, and node $n_1$ is the source node. The set of numbers associated with each node indicates the active time slots in the working period of that node. Fig.~\ref{fig:example}(b) and (c) are two multicast trees $T_1$ and $T_2$, respectively. If node $n_2$ in ${T_1}$ transmits at time
slot 3 or 4, then only one child node ($n_6$ or $n_4$) can receive the data.
Therefore, a correct transmission schedule of node $n_2$ for $T_1$ must be a
set of time slots which contains $\{3,4\}$.

From this observation, we introduce the concept of ``feasible schedule".
Actually, finding a feasible schedule of any non-leaf node in a multicast
tree equals to finding a ``hitting set"\cite{Garey1979}. We clarify this by Definition \ref{def:HittingSet}
and Definition \ref{def:FeasibleSchedule}.
\newtheorem{definition}{\textbf{Definition}}
\begin{definition}[Hitting Set\textup{\cite{Garey1979}}]
\label{def:HittingSet}
Given a collection $\cal C$ of subsets
of a finite set $\cal F$, a \textit{hitting set} is a subset $\cal {F}^{\prime } \subseteq \cal F$ such that $\cal {F}^{\prime }$
contains at least one element from each subset in $\cal C$.
\end{definition}
\begin{definition}[Feasible Schedule]
\label{def:FeasibleSchedule}
Given a multicast tree $T$ in $G$, a function $B:\mathit{nl}(T) \to {2^{\{ 1,2,...K\} }}$ is called a \textit{feasible
schedule} for $T$ if and only if for any $u \in \mathit{nl}(T)$, $B(u)$ is a hitting
set of the collection $\{ \Gamma (v)|v \in \mathit{child}(u,T)\}$.
\end{definition}

According to Definition \ref{def:FeasibleSchedule}, the energy cost for sending a data packet using a
feasible schedule $B$ on multicast tree $T$ can be written as $%
\sum_{u \in \mathit{nl}(T)} {|B(u)| \cdot {e_s}}$. This is only part of the
total energy cost in a multicast session, because the energy consumption for
receiving data also needs to be taken into account. In our case, since each forwarding node $u$ in
$\mathit{nl}(T)$ knows the duty-cycling schedules of all its neighboring nodes, it
can always transmit at the time slots in $B(u)$ such that each node in $\mathit{child}(u,T)$ receives the same data packet only once. Therefore, the energy
cost for receiving a data packet using multicast tree $T$ can be written as $(|N(T)| - 1) \cdot {e_r}$.

Based on the above discussions, we introduce the Minimum-Energy Multicasting Tree Construction and Scheduling (MEMTCS) problem in
Definition \ref{def:MEMTCS}:
\begin{definition}[MEMTCS]
\label{def:MEMTCS}
Given a DCWSN $G$, a terminal set $M \subseteq V$, and a source node $s\in M$, the MEMTCS problem seeks a 2-tuple $\langle T_{\mathrm{opt}},B_{\mathrm{opt}}\rangle$ in which $T_{\mathrm{opt}}$ is a
multicast tree rooted at $s$ and ${B_{\mathrm{opt}}}$ is a feasible schedule for $
T_{\mathrm{opt}}$, such that the total energy cost
\[\Pi (T_{\mathrm{opt}},B_{\mathrm{opt}})=\sum_{u\in \mathit{nl}(T_{\mathrm{opt}})}|B_{\mathrm{opt}}(u)|\cdot e_{s}+\left(|N(T_{\mathrm{opt}})|-1\right)\cdot {e_{r}}\]
is minimized.
\end{definition}

For example, we can find two feasible schedules $B_1$ and ${B_2}$ for ${T_1}$ and ${T_2}$ in Fig.~\ref{fig:example}, respectively,
such that ${B_1}(n_1)=\{1\}$, ${B_1}(n_2)=\{3,4\}$, ${B_1}(n_4)=\{2,3\}$, and ${B_2}(n_1)={B_2}(n_3) = {B_2}(n_5) = \{ 3\}$, ${B_2}(n_8) = \{ 2\}$.
Suppose that ${e_s} = 10$ and ${e_r} = 2$. We can get $\Pi ({T_1},{B_1}) =60$ and $\Pi ({T_2},{B_2}) =50$. Actually, $\langle T_2,B_2\rangle$ is the optimal solution for $n_1$ to send data to $\{n_6,n_7,n_8\}$ in Fig.~\ref{fig:example}.

\section{Solving the MEMTCS Problem}

We first briefly evaluate the hardness of MEMTCS. We prove it is NP-hard by a reduction from the Minimum Hitting Set (MHS) problem \cite{Garey1979}, and we claim this in Theorem 1:

\newtheorem{theorem}{\textbf{Theorem}}
\begin{theorem}
\label{thm:NP_MEMTCS}
The MEMTCS problem is NP-hard.
\end{theorem}

The MHS problem was proved to be equivalent to the Minimum Set Cover (MSC)
problem\cite{Garey1979,Ausiello1980}. Moreover, Fiege \cite{Feige1998} has proved that, unless NP has quasi-polynomial
time algorithms, there does not exist a polynomial-time algorithm for the
MSC problem with performance ratio of $(1-o(1))$ln$n$, where $n$ is the size of the MSC problem. Therefore, with the proof of Theorem~\ref{thm:NP_MEMTCS}, we can easily get:

\newtheorem{corollary}{\textbf{Corollary}}
\begin{corollary}
Unless $\mathit{NP}\subseteq \mathit{DTIME}\left(n^{\mathcal{O}(\log\log n)}\right)$, there does not exist a polynomial-time approximation algorithm with
performance ratio of $(1-o(1))\ln\Delta$ for the MEMTCS problem, where $\Delta$ is the maximum node degree of network graph under consideration.
\end{corollary}

Next, we propose an approximation algorithm for the MEMTCS
problem. We first provide a brief overview of our algorithm in Section~\ref{sec:overview}, then
describe our methods in details in Section~\ref{sec:gtrans}--\ref{sec:approx1}.

\subsection{An Overview of the Proposed Algorithm} \label{sec:overview}
Our approximation algorithm consists of several steps.
Firstly, we use a graph transformation method to transform the original network graph $G$ into an
\textit{extended graph} $\widetilde G$ where the possible transmitting time slots of the nodes in $G$ are represented as \textit{satellite nodes},
and the nodes in $\widetilde G$ are connected in a particular way to facilitate the design of our approximation algorithm.
Secondly, we propose the concept of \textit{Minimum Satellite Bridge} (MSB) in $\widetilde G$ as well as an algorithm for finding an approximation MSB. The MSB is actually a special tree in $\widetilde G$ whose nodes can cover all the nodes in $M$. Finally, we map the approximation for MSB to a multicast tree in $G$ and a feasible schedule for the multicast tree, which serves as an approximate solution to the MEMTCS problem.

To find the approximation ratio of our algorithm, we propose another concept \textit{Minimum Isotropic Scattering Tree} (MIST), which is a special multicast tree ${T_I}$ in $G$ spanning the nodes in $M$. We prove that ${T_I}$ serves as a quantitative ``bridge'' between the number of nodes in a MSB and $\langle T_{\mathrm{opt}},B_{\mathrm{opt}}\rangle$. As a result, we obtain the approximation ratio of our algorithm.

\subsection{Graph Transformation} \label{sec:gtrans}

The first step of our approach is to transform the original network graph
into an \textit{extended graph}.
We introduce the concept of the extended graph in Definition \ref{def:ExtendedGraph}:

\begin{definition}[Extended Graph]
\label{def:ExtendedGraph}
The \textit{extended graph} of $G$ is an undirected graph $\widetilde G=(\widetilde V,\widetilde E)$, where $\widetilde V$ is the set of nodes and $\widetilde E$ is the set of edges. The nodes and edges in $\widetilde G$ are created by the following steps:
\renewcommand{\labelenumi}{(\roman{enumi})}
\begin{enumerate}
\item Initially, $\widetilde V=V$ and $\widetilde E=\emptyset$;
\item For each node $u\in V$ and each time slot $i \in \bigcup_{v \in
\mathit{nb}_G(u)} {\Gamma (v)}$, create a new node $\lambda (u,i)$ in $\widetilde V$. The node $\lambda (u,i)$ is called a \textit{satellite node} of $u$ on slot $i$, and
$u$ is called a \textit{nuclear node} of $\lambda (u,i)$. The set of all satellite nodes of $u$ is denoted by $\Psi(u)$;
\item For each node $u\in V$, create an undirected edge between each pair
of nodes in $\Psi(u)\cup\{u\}$. In other words, the sub-graph induced by $\Psi(u)\cup\{u\}$ is a complete graph;
\item For each edge $(u,v)\in E$, each time slot $i\in \Gamma(v)$ and each
time slot $j\in \Gamma(u)$, create three undirected edges $(\lambda
(u,i),\lambda (v,j))$, $(\lambda (u,i),v)$ and $(u,\lambda (v,j))$.
\end{enumerate}
\end{definition}

From Definition~\ref{def:ExtendedGraph}, we can see that $\widetilde V$ can be partitioned into
two disjoint subsets: $V$ and ${V_S}$, where $V$ is the set of all nuclear
nodes, and ${V_S}$ is the set of all satellite nodes. A nuclear node may
have multiple satellite nodes, but any satellite node only has one nuclear
node. An example of the extended graph is shown in Fig.2.
\begin{figure}[htbp]
\centerline{\includegraphics{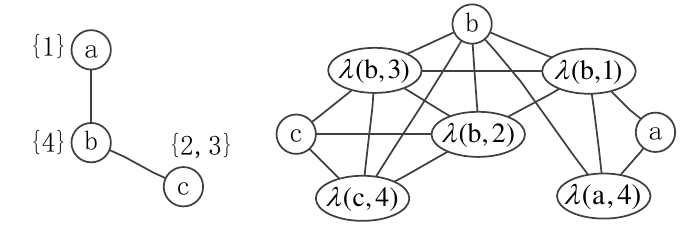}}
\label{ExtendGraph}
\centering
\caption{A DC-WSN graph $G$ (left) and its extended graph $\widetilde G$ (right).}
\end{figure}

According to the construction rules in Definition~\ref{def:ExtendedGraph}, we can obtain some useful properties of $\widetilde G$, as described in Lemma~\ref{lma:extf1} and Lemma~\ref{lma:extf2}. Their proofs are omitted due to page limits.
\begin{lemma} \label{lma:extf1}
There are at most $(K+1)|V|$ nodes and $\left(
\begin{array}{c}
K+1 \\
2%
\end{array}%
\right) |V|+3K^{2}|E|$ edges in $\widetilde G$.
\end{lemma}

\begin{lemma} \label{lma:extf2}
Any two nuclear nodes cannot be adjacent in $\widetilde G$, and any
satellite node in $V_{S}$ can be adjacent to at most $\Delta +1$ nuclear nodes in $\widetilde G$.
\end{lemma}

\subsection{Minimum Satellite Bridge} \label{sec:minsb}

We first introduce the concept of Minimum Satellite Bridge (MSB) in Definition~\ref{def:MSB}, and then we propose an approximate algorithm for finding a MSB.  As we will see later, finding a MSB is an important building block of our algorithm for solving the MEMTCS problem.

\begin{definition}[Minimum Satellite Bridge]
\label{def:MSB}
Given $\widetilde G = (\widetilde V,\widetilde E)$ and the terminal set $M \subseteq V$, a \textit{Satellite Bridge} $\mathit{SB}$ is a sub-tree of $\widetilde G$  which satisfies:
\renewcommand{\labelenumi}{\arabic{enumi})}
 \begin{enumerate}
\item The nodes in $SB$ are all satellite nodes;
\item Each node in the terminal set $M$ is adjacent to at least one node in $SB$.
\end{enumerate}
The \textit{Minimum Satellite Bridge} (MSB) $SB^\ast$ is an $\mathit{SB}$ with a minimum number of nodes.
\end{definition}

Next, we propose an approximation algorithm with a performance ratio of $\mathcal{O}\left(H(\Delta+1)\right)$ for finding a MSB, as shown in Algorithm~\ref{alg:MSB}.
%
\begin{algorithm}[htb]
   \caption{Finding an approximate MSB}
   \textbf{Input:} The extended graph $\widetilde{G}$ and the terminal set $M$.\\
   \textbf{Output:} An approximate MSB $\overline{\mathit{SB}}$.
   \begin{algorithmic}[1]
    \State{$C = \emptyset, \mathit{UC} = M$;}
    \While {$\mathit{UC} \ne \emptyset$}
    \State{$v = \arg\max_{u \in V_S - C} \left|\mathit{nb}_{\widetilde{G}} (u) \cap \mathit{UC} \right|$;}
    \State{$C = C \cup \{v\}$;}
    \State{$\mathit{UC} = \mathit{UC} - (\mathit{nb}_{\widetilde{G}} (v) \cap \mathit{UC})$;}
    \EndWhile
    \State{Let ${\widetilde G_s}$ be the sub-graph of ${\widetilde G}$ induced by $V_S$. Assign each edge in $\widetilde{G}_s$ a weight of 1. Compute an approximate minimum Steiner tree $\mathit{ST}$ in $\widetilde{G}_s$ which connects the nodes in $C$;}
    \State{\Return $\overline{\mathit{SB}} = \mathit{ST}$.}
   \end{algorithmic}
   \label{alg:MSB}
\end{algorithm}
This algorithm consists of two stages. The first stage is lines 1-6, and the second stage is line 7. In the first stage, we use a greedy set cover algorithm to find a small satellite node set $C$ that can cover all the nodes in $M$. In each loop, we first find a node $v$ which has the maximum number of adjacent nodes in the uncovered node set $\mathit{UC}$ (line 3). Then we add $v$ into $C$ (line 4) and update $\mathit{UC}$ (line 5). In the second stage (line 7), an approximate Steiner tree algorithm is applied upon $C$.

Since each satellite node can cover at most $\Delta$+1 nodes in $M$ (according to Lemma~\ref{lma:extf2}), the first stage of Algorithm 1 has an approximation ratio of $H(\Delta +1)$ \cite{Cormen2001}. The approximation ratio of the second stage is a constant $\rho$. By taking advantage of the special structure of the extended graph, we can get:

\begin{theorem} \label{thm:MSB}
The approximation ratio of Algorithm~\ref{alg:MSB} is $3\rho H(\Delta  + 1) + \rho$, where $H(\cdot)$ is the harmonic number and $\rho$ is the approximation ratio of the algorithm used in Algorithm 1 for finding a minimum Steiner tree.
\end{theorem}

\subsection{Minimum Isotropic Scattering Tree} \label{sec:minist}
Now we show that an MSB can be mapped to a special multicast tree in $G$ called the \textit{Minimum Isotropic Scattering Tree} (MIST) and a feasible schedule for the internal nodes in the MIST. Although MSB does not produce an optimal solution to the MEMTCS problem, we show that it leads to an approximation for MEMTCS, and we use MIST as a medium to derive the approximation ratio. We first introduce the concept of MIST in Definition~\ref{def:MIST}:

\begin{definition}[Minimum Isotropic Scattering Tree]
For any tree $T$ in $G$ and any node $u \in {d^ + }(T)$, we denote by $\Upsilon (u,T)$ a minimum hitting set of the collection $\{ \Gamma (v)  |  v \in n{b_T}(u)\}$, and define $\Xi (T)=\sum_{u  \in {d^ + }( T)} \left|\Upsilon (u,T)\right|$. The \textit{Minimum Isotropic Scattering Tree} (MIST) $T_I$ is a tree in $G$ such that $T_I$ spans $M$ and that $\Xi (T_I)$ is minimized. \label{def:MIST}
\end{definition}

We claim that $\Xi (T_I )$ actually equals to the number of nodes in a
minimum satellite bridge, which is proved by Lemma~\ref{lma:ubd}, Lemma~\ref{lma:lbd}, and Theorem~\ref{thm:eq}.
Lemma~\ref{lma:ubd} claims that the number of nodes in a MSB is no more than $\Xi (T_I )$, whereas Lemma~\ref{lma:lbd} implies that $\Xi (T_I )$ is no more than the number of nodes in a MSB. With these two lemmas, Theorem 3 can be readily proved.

\begin{lemma} \label{lma:ubd}
$\vert N(\mathit{SB}^{\ast})\vert \le \Xi (T_I )$.
\end{lemma}

\begin{lemma} \label{lma:lbd}
Any satellite bridge $\mathit{SB}$ can be mapped to a 2-tuple $\langle R,F\rangle$, where $R$ is a tree in $G$ spanning the nodes in $M$, and $F$ is a function that satisfies:
\begin{enumerate}
\item For any $u \in d^ + (R), F(u)$ is a hitting set of the collection $\{\Gamma (v)  | v \in \mathit{nb}_R (u)\}$;
\item $\sum_{u \in d^ + (R)} |F(u)| \le | N(\mathit{SB})|$.
\end{enumerate}
\end{lemma}

\begin{theorem} \label{thm:eq}
$\Xi (T_I ) = | N(\mathit{SB}^\ast)|$
\end{theorem}

The proofs take advantage of the special structure of the extended graph. In addition, the proof of Lemma~\ref{lma:lbd} actually provides a method of mapping any satellite bridge to a multicast tree and a feasible schedule for the internal nodes in the multicast tree. This mapping can be roughly described as follows. According to the construction rules of the extended graph, the satellite nodes in a satellite bridge can be mapped to the nuclear nodes that they belong to, as well as the transmitting time slots on these nuclear nodes. Furthermore, we can find a multicast tree spanning these nuclear nodes and the destination nodes in $M$, and the internal nodes in the multicast tree are all the mapped nuclear nodes. According to the special node-connecting method of the extended graph, the mapped transmitting time slots of any internal node in the multicast tree can cover all its neighboring nodes in the tree.

\subsection{Approximation Algorithm for MEMTCS} \label{sec:approx1}

Based on the methods introduced by the previous sections, we propose our algorithm for the MEMTCS problem, as shown in Algorithm~\ref{alg:MEMTCS}.
\begin{algorithm}[htb]
   \caption{Approximation for MEMTCS}
   \textbf{Input:} A DC-WSN $G$, a terminal set $M$, and a source node $s \in M$. \\
   \textbf{Output:} A multicast tree $\overline{T}$ and a feasible schedule $\overline{B}$.
   \begin{algorithmic}[1]
    \State{Construct the extended graph $\widetilde{G} = (\widetilde{V},\widetilde{E})$ of $G$;}
    \State{Use Algorithm~\ref{alg:MSB} to compute an approximate minimum satellite bridge $\overline{\mathit{SB}}$;}
    \State{Use the method in the proof of Lemma 4 to map $\overline{\mathit{SB}}$ to a 2-tuple $\langle \widehat{T},\widehat{F} \rangle$. Let $\overline{T}$ be the rooted tree got by designating $s$ as the root of $\widehat{T}$;}
    \For{each node $u \in \mathit{nl}(\overline{T})$}
        \If {$u \in d^ + (\overline{T})$}
            \State{$\overline{B}(u) = \widehat{F}(u)$;}
        \Else
            \State{Let $v$ be $u$'s child node in $\overline{T}$. Find an arbitrary} \State{$i \in \Gamma (v)$ and let $\overline{B}(u) = \{i\}$;}
        \EndIf
    \EndFor
    \State{\Return $\langle \overline{T},\overline{B} \rangle$.}
   \end{algorithmic}
   \label{alg:MEMTCS}
\end{algorithm}

The output of Algorithm~\ref{alg:MEMTCS} is a 2-tuple $\langle \overline{T},\overline{B} \rangle$. From Lemma 4, It is easy to know that $\overline{T}$ is a multicast tree spanning the nodes in $M$ and $\overline{B}$ is a feasible schedule for $\overline{T}$. Next, we prove the approximation ratio of Algorithm~\ref{alg:MEMTCS} by Lemma~\ref{lma:optp1}, Lemma~\ref{lma:optp2} and Theorem~\ref{thm:approxr1}. Lemma~\ref{lma:optp1} is actually based on a special property of the MHS problem, i.e, if we add a subset in an instance of the MHS problem, then the cardinality of the result minimum hitting set will increase at most 1. Using Lemma~\ref{lma:optp1}, Lemma~\ref{lma:optp2} finds out a quantitative relationship between $\Xi (T_I)$ and $\langle T_{\mathrm{opt}},B_{\mathrm{opt}}\rangle$, which is used in the proof of Theorem~\ref{thm:approxr1}.

\begin{lemma} \label{lma:optp1}
For any node $u \in d^+ (T_{\mathrm{opt}})$, we have \[| \Upsilon (u,T_{\mathrm{opt}} )| \le | B_{\mathrm{opt}} (u)|+1.\]
\end{lemma}

\begin{lemma} \label{lma:optp2}
$\langle T_{\mathrm{opt}},B_{\mathrm{opt}}\rangle$ is related to $T_I$ by \[\textstyle{\sum_{u \in nl(T_{\mathrm{opt}})}} |B_{\mathrm{opt}}(u)| \ge \Xi (T_I) - |d^+ (T_{\mathrm{opt}})| + 1.\]
\end{lemma}

\begin{theorem} \label{thm:approxr1}
The approximation ratio of Algorithm~\ref{alg:MEMTCS} is $12\rho H(\Delta + 1) + 4\rho$.
\end{theorem}

The dominating running time of Algorithm~\ref{alg:MEMTCS} is the time on constructing $\overline{\mathit{SB}}$ in line 2, using Algorithm~\ref{alg:MSB}. Lines 1-6 in Algorithm~\ref{alg:MSB} can be implemented in $\mathcal{O}(|\widetilde{V}|^2)$ time. If we use the 2-approximation algorithm proposed in \cite{Mehlhorn1988} to compute an approximate minimum Steiner tree in line 7, the resulting time complexity is $\mathrm{O}(|\widetilde{V}|\log |\widetilde{V}| + |\widetilde{E}|)$. Given $K$ as a predefined constant, the time complexity of Algorithm~\ref{alg:MEMTCS} is $O(|V|^2 + |E|)$, and the
approximation ratio is $24H(\Delta + 1) + 8$.

\section{Distributed Implementation}

In this section, we provide a distributed implementation of Algorithm~\ref{alg:MEMTCS} for the MEMTCS problem. Note that the main operation of Algorithm~\ref{alg:MEMTCS} is line 2, in which Algorithm~\ref{alg:MSB} is called to compute an approximate minimum
satellite bridge. Therefore, we first propose the distributed implementation
of Algorithm~\ref{alg:MSB}.

As we have described in Section~\ref{sec:minsb}, Algorithm~\ref{alg:MSB} consists of two stages: the first stage is lines 1-6, in which a greedy strategy is used to find a small satellite node set $C$ covering the nodes in $M$, and the second stage is line 7, in which an approximate minimum Steiner tree is computed. The first stage of Algorithm~\ref{alg:MSB} can be decentralized in a way similar with the distributed dominating set algorithm in \cite{Liang2000}, which is shown in Algorithm~\ref{alg:Distributed}.
\begin{algorithm}[htb]
   \caption{Distributed implementation of the first stage of Algorithm~\ref{alg:MSB}}
   \begin{algorithmic}[1]
    \State{Each white node $u$ with non-empty $\mathit{rnb}(u)$ broadcasts a message (``election'', $|\mathit{rnb}(u)|$, $u$.ID);}
    \State{Each red node checks all the ``election'' messages it receives, and finds a node $v$ whose value of $|\mathit{rnb}(v)|$ is the maximum (break tie by choosing the node with largest ID). Then it sends $v$ a message ``you win'';}
    \State{If a white node $u$ receives ``you win'' messages from all nodes in $\mathit{rnb}(u)$, then it colors itself blue, and broadcasts a message ``I am dominator'';}
    \State{If a red node receives an ``I am dominator'' message, then it colors itself green, and broadcasts a message ``I am dominated'';}
    \State{If a white node $u$ receives a ``I am dominated'' message from a neighboring node $w$, then it deletes $w$ from \textit{rnb}($u)$.}
   \end{algorithmic}
   \label{alg:Distributed}
\end{algorithm}

In Algorithm~\ref{alg:Distributed}, each node in $\widetilde{G}$ is colored red, green, white, or blue. The red nodes are the nodes in $M$ which are not covered yet. The green nodes are the nodes in $M$ which are already covered by some blue nodes. The blue nodes are the nodes which are selected into the resulting node set $C$, and the white nodes are the nodes which are not selected. Initially, each node in $M$ is colored red, and all other nodes are colored white. Besides, each white node $u$ owns a set $\mathit{rnb}(u)$, which is initialized to be the set of IDs of all $u$'s red neighbors.

It can be seen that Algorithm~\ref{alg:Distributed} is a faithful implementation of the greedy strategy in lines 1-6 of Algorithm~\ref{alg:MSB}, so the approximation ratio of Algorithm~\ref{alg:Distributed} is $H(\Delta + 1)$. There are at most $|M|$ rounds in Algorithm~\ref{alg:Distributed}, because it repeats until no red nodes exist, and at least one red node
turns green in each round. Therefore, the message complexity of Algorithm~\ref{alg:Distributed} is $\mathcal{O}(|M|\cdot|\widetilde{V}|) = \mathcal{O}(|M|\cdot|V|)$.

The second stage of Algorithm~\ref{alg:MSB} can be decentralized by using a distributed Steiner tree algorithm in the literature \cite{Bauer1996,Chalermsook2005,Muhammad2006}.
If we adopt the 2-approximation distributed algorithm proposed in
\cite{Bauer1996}, then the message complexity is $\mathcal{O}(|M|\cdot|V|)$ and the time complexity is $\mathcal{O}(|M|\cdot D)$,
where $D$ is the diameter of $G$.

The distributed implementation of Algorithm~\ref{alg:MEMTCS} (except line 2) is trivial: an arbitrary spanning tree of the sub-graph induced
by $N(\overline{\mathit{SB}})$ needs to be found in line 3. To accomplish this, the distributed \textit{Depth-First Search} (DFS) algorithm proposed by Makki
\textit{et al}.\cite{Makki1996} can be applied. The time complexity and
message complexity of the distributed DFS algorithm are both $\mathcal{O}(|V|)$.

Based on these discussions, we can get Corollary 2:
\begin{corollary}
There exists a distributed algorithm for MEMTCS. It has an approximation ratio of $24H(\Delta + 1) + 8$. The time complexity and message complexity of the distributed algorithm are $\mathcal{O}(D\cdot|V|)$ and $\mathcal{O}(|M|\cdot|V|)$, respectively, where $D$ is the diameter of $G$.
\end{corollary}

Though our distributed implementation for Algorithm~\ref{alg:MEMTCS} is based on the extended graph $\widetilde{G}$, it can be easily adapted to run on $G$. According to the construction method of $\widetilde{G}$, any satellite node in $V_S$ can be seen as a local ``pseudo node'' administrated by its nuclear node in $V$. Therefore, each nuclear node can send messages for its satellite nodes, and do the computation that its satellite nodes need to do. Let $\widetilde{\cal A}$ be the distributed algorithm running on $\widetilde{G}$, and let $\cal A$ be the adapted version of $\widetilde{\cal A}$ running on $G$. If several satellite nodes of the same nuclear node $u$ send their messages simultaneously to other nuclear nodes in $\widetilde{\cal A}$, then $u$ in $\cal{A}$ can send these messages at different time slots. Note that any nuclear node has at most $K$ satellite nodes and $K$ is a predefined constant. Therefore, if we define a constant $\delta$($\delta>K$), then any nuclear node $u$ in $\cal A$ can receive messages for all its satellite nodes after waiting $\delta$ time, and then $u$ can do the computation for its satellite nodes based on the messages it receives. The message passing processes between the satellite nodes administrated by the same nuclear node in $\widetilde{\cal A}$ become local computations of that nuclear node in $\cal A$, and no new messages are generated in $\cal A$. Therefore, $\cal A$ has the same time and message complexity as $\widetilde{\cal A}$.

\begin{figure*}[htbp]
\centering
\subfigure[]{
\includegraphics[scale=0.36]{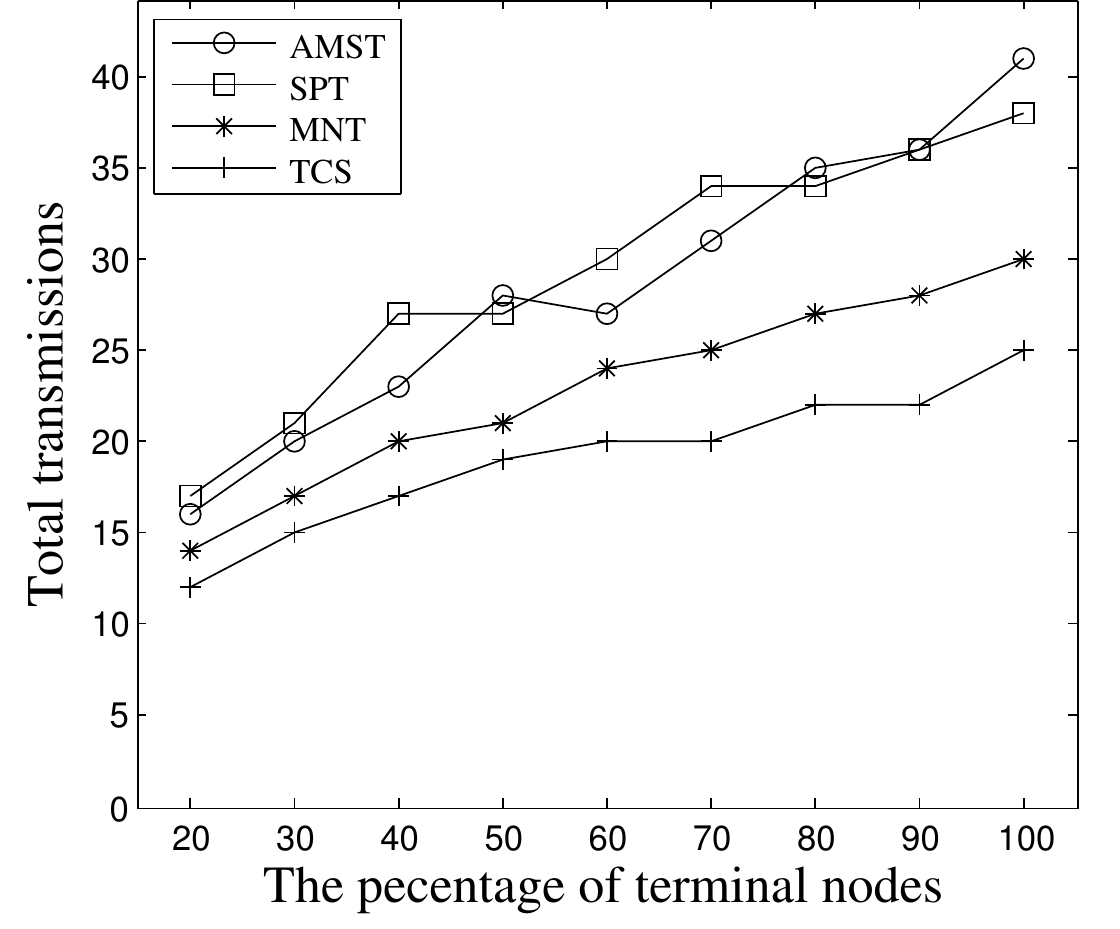}
\label{fig:TCSsubfig1}
}
\subfigure[]{
\includegraphics[scale=0.36]{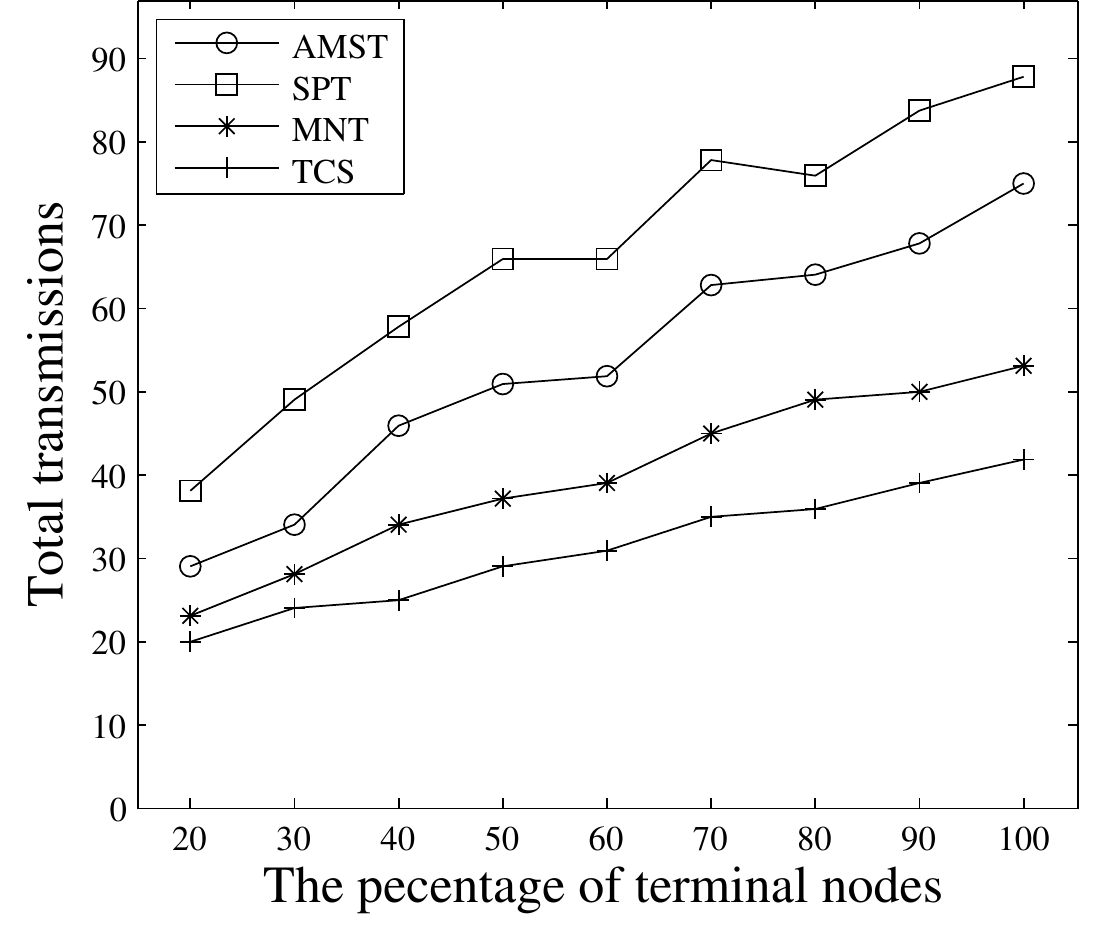}
\label{fig:TCSsubfig2}
}
\subfigure[]{
\includegraphics[scale=0.36]{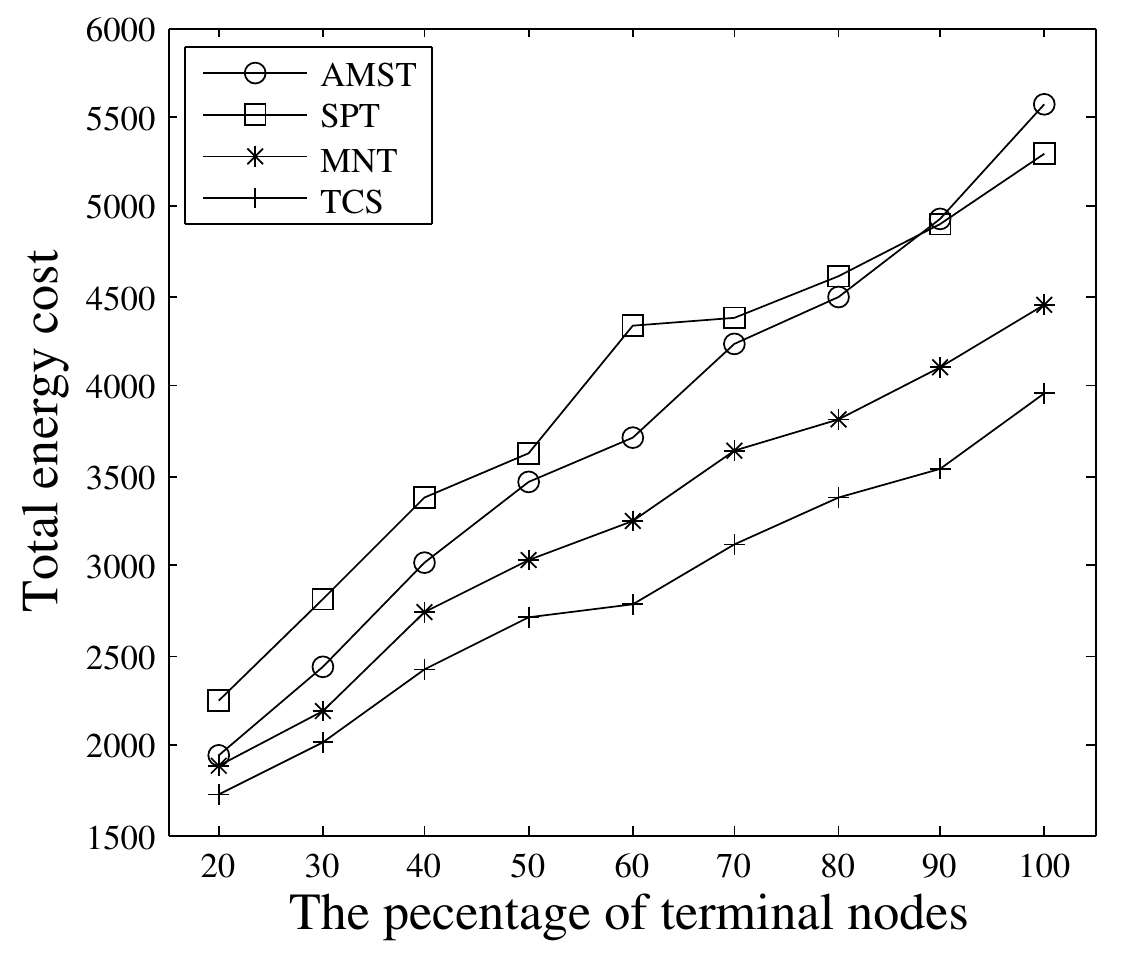}
\label{fig:TCSsubfig3}
}
\subfigure[]{
\includegraphics[scale=0.36]{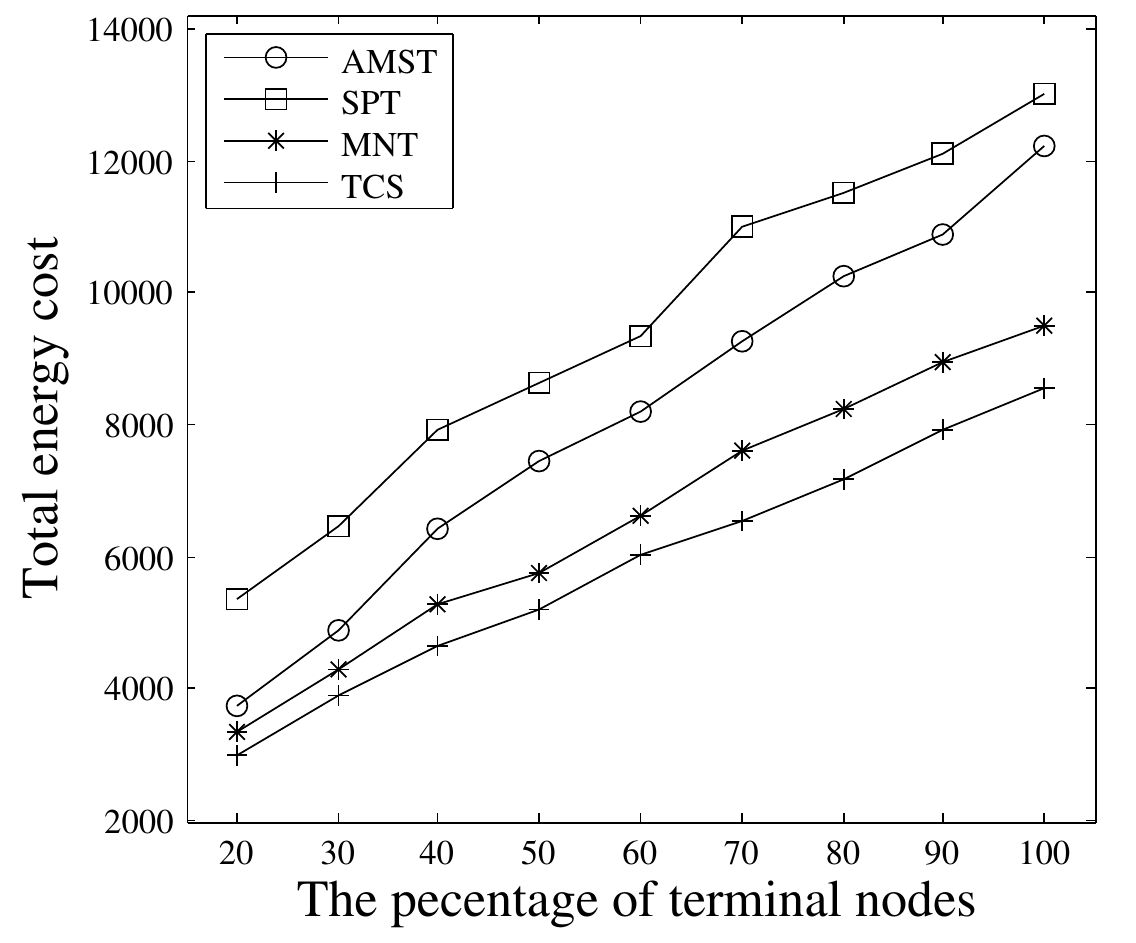}
\label{fig:TCSsubfig4}
}
\caption[Optional caption for list of figures]{ Performance evaluation of various multicasting algorithms. The percentage of terminal nodes scales from 20\% to 100\%. \subref{fig:TCSsubfig1} $\vert V\vert$=100, \subref{fig:TCSsubfig2} $\vert V\vert$=300, \subref{fig:TCSsubfig3} $\vert V\vert$=100, \subref{fig:TCSsubfig4} $\vert V\vert$=300.} \label{fig:TCS}
\end{figure*}

\section{Performance Evaluation}

In this section we evaluate the performance of our algorithm via
simulations. The simulations focus on the effect of various network
conditions on the performance of various multicasting
algorithms. In the simulations, we deploy wireless nodes randomly
in a $1000m\times 1000m$ square, and the transmission range of each node is
set to 300 meters. Each node randomly picks some time slots in the working
period as its active time slots. Without loss of generality, the energy cost
for sending and receiving a data packet by any node is set to 100 and 15,
respectively.

To the best of our knowledge, there is no polynomial-time minimum-energy
multicasting algorithms designed for DC-WSNs under a generic duty-cycling
model. So we have to compare our algorithms with several traditional
multicasting algorithms, including the \textit{Shortest Path Tree} (SPT) algorithm, the \textit{Approximate Minimum Steiner Tree} (AMST) algorithm, and the minimal data overhead tree (the MNT algorithm) proposed by \cite{Ruiz2005,Ruiz2007}. The SPT algorithm computes shortest paths from the source node to the receiver nodes, and
aggregates these shortest paths to construct a multicast tree. The AMST
algorithm computes an approximate minimum Steiner tree spanning all the
nodes in the terminal set $M$. Here, we adopt the AMST algorithm provided by
Kou \textit{et al}.\ \cite{Kou1981}, which was also used by Liang \textit{et al}.\ \cite{Liang2009}
to solve the minimum-energy all-to-all multicasting problem in always-active
wireless ad-hoc networks. The MNT algorithm was designed for reducing the multicast redundancy in static wireless ad-hoc networks. The work in \cite{Ruiz2005} and \cite{Ruiz2007} has proved that MNT can reduce the redundant transmissions in a multicast session more effectively than other heuristics.\footnote{We refrain from comparing with the algorithm proposed in \cite{Liang2006} and oCast \cite{Su2009}, as they both have a much higher complexity than ours: $\mathcal{O}(|M|^2\cdot|V|^2)$ for the former and exponential in $|M|$ for the latter.}

To use the multicast trees constructed by the SPT, AMST and MNT algorithms
in a DC-WSN environment, we need to find a transmission schedule for each
forwarding node in the multicast trees. Obviously, the most energy-efficient
transmission schedule for any forwarding node $u$ in a multicast tree $T$ is the
minimum hitting set of the collection $\{\Gamma (v) | v \in \mathit{child}(u,T)\}$. However, finding a minimum hitting set is a NP-hard problem.
Therefore, we use a greedy hitting set algorithm\cite{Johnson1973} to find the transmission schedules of
the forwarding nodes. Since the greedy algorithm is essentially the
best-possible polynomial-time approximation algorithm for the minimum
hitting set problem (unless $\mathit{NP} \subseteq \mathit{DTIME}(n^{\mathcal{O}(\log \log n)}))$,
each forwarding node in the multicast trees generated by the SPT, AMST and MNT algorithms has the best-possible energy-efficient transmission
schedule.

In the simulations, we compare our algorithms with the traditional ones
using two metrics including the total energy cost and the transmission
redundancy. Note that minimizing the transmission redundancy is an important
optimization objective both in multicasting and in broadcasting\cite{Hong2010,Su2009,Wang2009,Guo2009}.
However, to the best of our knowledge, there is no other work that provides polynomial-time
algorithms with guaranteed approximation ratios for minimizing the
transmission redundancy of multicasting/broadcasting in DC-WSNs under a generic duty-cycling
model.

In Fig.~\ref{fig:TCS}, we compare Algorithm~\ref{alg:MEMTCS} (denoted by TCS) with SPT, AMST and
MNT. The length of working period is
set to 20, and the percentage of terminal nodes scales from 20{\%} to 100{\%}
with an increment of 10{\%}. The algorithm proposed in \cite{Kou1981} is adopted in Algorithm 2 for computing an
approximate minimum Steiner tree. Fig.~\ref{fig:TCS}(a) and (b) plot the number of total transmissions in a multicast session. The number of network nodes is set to 100 and 300,
respectively. We can see that MNT outperforms SPT and AMST greatly, because the multicast tree generated by MNT has less forwarding nodes (non-leaf nodes) than the other multicast trees\cite{Ruiz2005}. We also see that TCS significantly outperform the other
algorithms, and the transmission redundancy is reduced by about 20{\%} as
the percentage of terminal nodes approaches 90{\%}. The reason is that,
since the traditional SPT, AMST and MNT algorithms generate multicast trees
regardless of the duty cycles of the wireless nodes, they can't optimize the
transmission schedules of the forwarding nodes in a global manner. Therefore,
although the transmission schedules are optimized locally in the SPT, MST and
MNT algorithms using a best-possible optimization algorithm (the greedy
algorithm), their transmission redundancies are still high. On the contrary,
the TCS algorithm builds the multicast tree and finds the transmission
schedules of the forwarding nodes in a holistic manner by taking advantage of
the special structure of the extended graph, so the transmission redundancy is
reduced more effectively than the traditional algorithms.
\begin{figure}[htp]
\subfigure[]{
\includegraphics[scale=0.345]{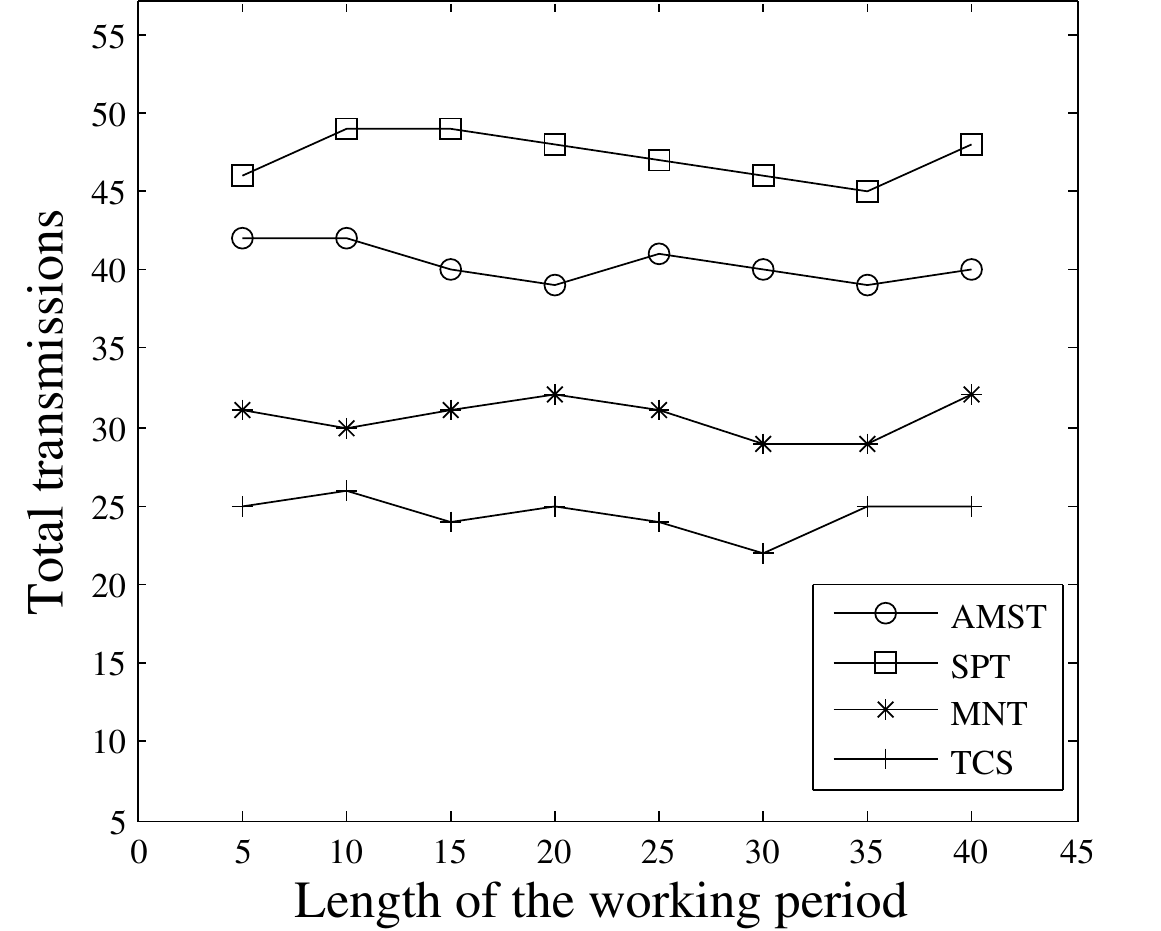}
\label{fig:WPsubfig1}
}
\subfigure[]{
\includegraphics[scale=0.345]{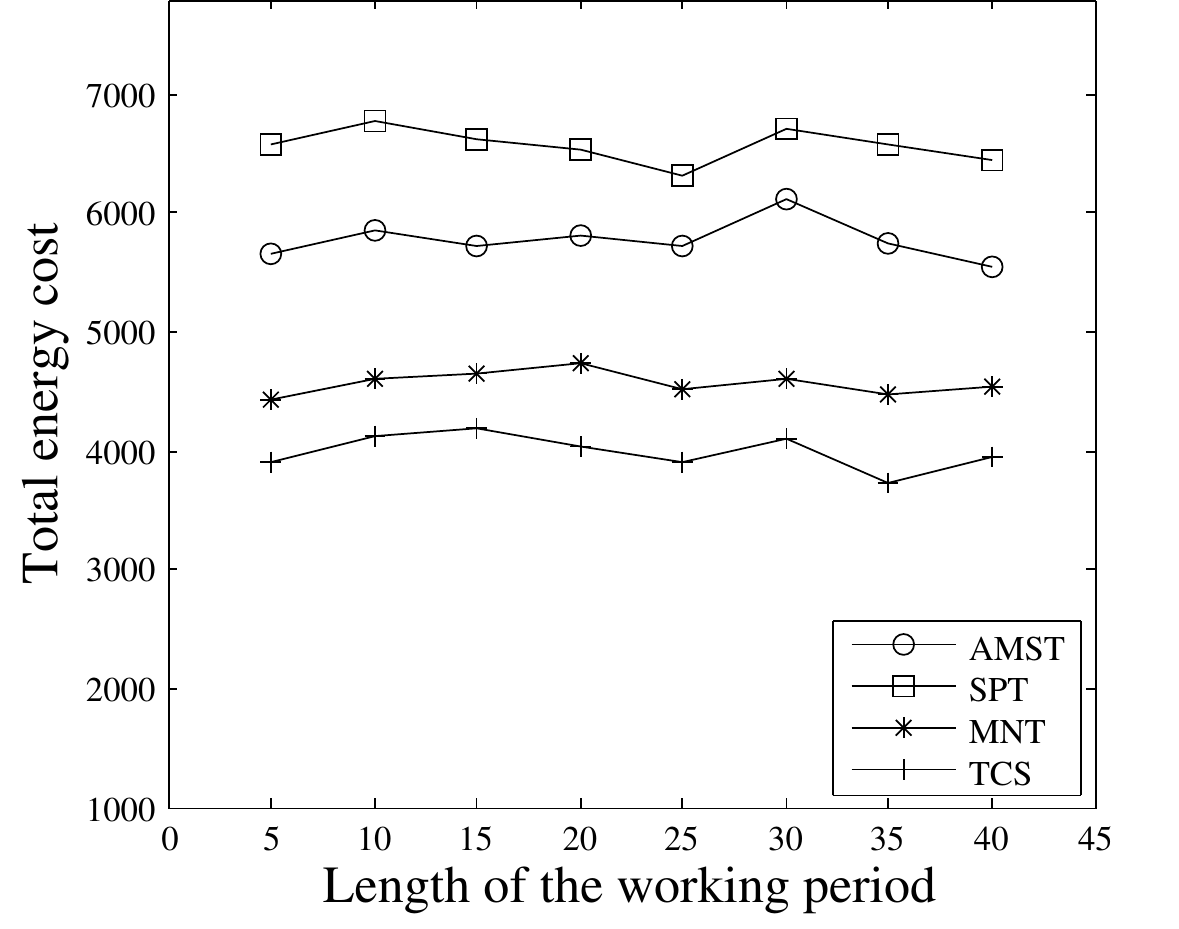}
\label{fig:WPsubfig2}
}
\caption[Optional caption for list of figures]{ Performance evaluation of various multicasting algorithms under scenarios with different lengths of the working period. The length of the working period scales from 5 to 40. The number of nodes in $V$ is 150 and the number of nodes in $M$ is 100.} \label{fig:WP}
\end{figure}

Fig.~\ref{fig:TCS}(c) and (d) plot the total energy costs of different multicasting algorithms. Again, the numbers of network nodes in Fig.3(c) and (d) are set to
100 and 300, respectively. We can see that the TCS algorithm still outperforms the other algorithms.
The explanation is that, compared with the other algorithms, the TCS algorithm significantly reduces the number of total transmissions, and generates a multicast tree with moderate number of nodes. As a result, the total energy cost of the TCS algorithm is lower than the other algorithms when both the energy cost for sending data and the energy cost for receiving data are considered.
We also notice that the TCS algorithm performs better when the percentage of destination nodes increases. This is because that more redundant transmissions are reduced by TCS when the multicast tree grows.

In Fig.~\ref{fig:WP} we study how the length of the working period impacts the performance of our algorithms. The number of network nodes is set to 150, and the number of terminal nodes is set to 100. The length of the working period scales from 5 to 40 with an increment of 5. Fig.~\ref{fig:WP}(a) and (b) plot the number of total transmissions and the total energy cost for multicasting, respectively. It is evident that our algorithm still outperform the other algorithms when the length of the working period changes. Meanwhile, we also see that, for all algorithms compared in Fig.~\ref{fig:WP}(a)-(b), both the transmission redundancy and the total energy cost do not vary much. An explanation is that, since the active time slots of any node are randomly selected from the working period, the number of common active time slots of any two different nodes does not vary much when the length of the working period increases. Therefore, the performance of all these algorithms are rather independent of the length of the working period under our generic duty-cycling model.

\section{Conclusion}
In this paper, we have studied the Minimum-Energy Multicasting (MEM) problem in Duty-Cycled Wireless Sensor Networks (DC-WSNs). We formalized the Minimum-Energy Multicasting Tree
Construction and Scheduling (MEMTCS) problem, and proved its NP-hardness. A lower bound on the approximation ratio of any polynomial-time
algorithm for the MEMTCS problem was given in our work. We
presented an approximation algorithm with guaranteed approximation ratio for
the MEMTCS problem, and proposed a distributed
implementation of our algorithm. The simulation results demonstrate that
our algorithm outperform other related algorithms in terms of both the
total energy cost and the transmission redundancy.


\appendix[Proofs for the MEMTCS problem]
\begin{IEEEproof}[\textbf{Proof of Theorem 1}]
Given an instance ($\cal C,\cal F$) of the minimum hitting set problem, we create a
wireless network graph $G$ by the following method:

Let the elements in $\cal F$  be {\{}$f_1 ,f_2 ,...,f_p ${\}}, and let the subsets
in $\cal C$ be $C_1 ,C_2 ,...,C_q $. For each $C_j (1 \le j \le q)$, create a node
$v_j $ in $G$, and let $\Gamma (v_j )=\{i\vert (1 \le i \le p)
\wedge (f_i \in C_j )\}$. Create a node $x$ in $G$ , and connect $x$ to each $v_j (1
\le j \le q)$.

Let $x$ be the source node and $\{x\} \cup \{v_j \vert 1 \le j \le q\}_{ }$be
the terminal set $M$ in multicasting. Let $e_s = 1$ and $e_r = 0$. It is easy
to prove that $\mathcal{F}$ has a hitting set for $\mathcal{C}$ of size at most $k$ if and only if $G$ has a
multicast tree $T$ and a feasible schedule $B$ for $T$ such that $\Pi (T,B) \le k$.
Therefore, the MEMTCS problem is NP-hard.
\end{IEEEproof}

%

\begin{IEEEproof}[\textbf{Proof of Theorem 2}]
Each node $u \in {V_S}$ may cover a set $n{b_{\widetilde G}}(u) \cap M$, whose cardinality is bounded by $\Delta+1$ (see Lemma 2). Therefore, lines 1-6 of Algorithm~\ref{alg:MSB} intrinsically represent a greedy algorithm for finding a minimum set cover\cite{Cormen2001}. Suppose that $C^\ast$ is a smallest set of satellite nodes covering the nodes in $M$, we have:
\begin{equation}
|C|\le H(\Delta  + 1)|C^{\ast }| \label{eq:scb1}
\end{equation}
According to Definition~\ref{def:MSB}, we know that the nodes in $\mathit{SB}^\ast$ also cover the nodes in $M$. Hence
\begin{equation}
|C^{\ast }| \le |N(\mathit{SB}^{\ast })|
\end{equation}
Let $\mathit{ST}^\ast$ be a minimum Steiner tree in $\widetilde{G}_s$ which connects the nodes in $C$.  We have:
\begin{equation}
|E(\overline{\mathit{SB}})| \le \rho |E(\mathit{ST}^\ast)|
\end{equation}

For any node $c \in C - N(\mathit{SB}^{\ast})$, there must exist a node $m \in M$ adjacent to $c$.  Since $SB^\ast$ is a satellite bridge, $m$ must be adjacent to a certain node $t$ in $SB^\ast$. For convenience, we assume that $c \notin \Psi (m)$ and $t \notin \Psi (m)$ (otherwise the theorem can be proved in a similar way). According to Definition \ref{def:ExtendedGraph}(iv), we know that there must exist two nodes $c',t' \in \Psi (m)$ such that $c'$ is adjacent to $c$, and   $t'$ is adjacent to $t$. Since $c'$ and $t'$ are both satellite nodes of $m$, if $c' \ne t'$, then $c'$ and $t'$ must be adjacent, according to Definition \ref{def:ExtendedGraph}(iii). Therefore, there exists a path from $c$ to $t$ in $\widetilde{G}_s$ whose length is no more than 3, as shown in Fig.~\ref{fig:edge}.
In other words, there must exist a tree in ${\widetilde G_s}$ whose node set
\begin{figure}[htbp]
\centerline{\includegraphics[scale=0.8]{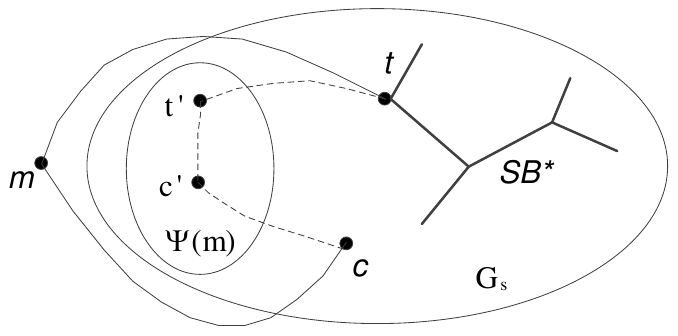}}
\centering
\caption{Connecting the node $c$ in $C$ to $\mathit{SB}^\ast$}
\label{fig:edge}
\end{figure}
contains $N(SB^{\ast }) \cup C$, and it has at most $|N(SB^{\ast })| + 3|C| - 1$ edges. Since $\mathit{ST}^\ast$ is a minimum Steiner tree that connects the nodes in $C$, it follows that:
\begin{equation}
|E(\mathit{ST}^\ast)| \le |N(\mathit{SB}^\ast)| + 3|C| - 1 \label{eq:edgeb}
\end{equation}
Finally, with equation (\ref{eq:scb1})--(\ref{eq:edgeb}), we can get:
\begin{eqnarray}
|N(\overline{\mathit{SB}})| &=& |E(\overline{\mathit{SB}})| + 1 ~~\le~~ \rho |E(\mathit{ST}^\ast)| + 1 \nonumber \\
&\le& \rho (|N(\mathit{SB}^\ast)| + 3|C| - 1) + 1 \nonumber \\
&\le& \rho (|N(\mathit{SB}^\ast)| + 3H(\Delta +1)|C^\ast| - 1) + 1 \nonumber\\
&\le& (3\rho H(\Delta +1)+\rho )|N(\mathit{SB}^\ast)| - (\rho - 1) \nonumber\\
&\le& (3\rho H(\Delta +1)+\rho )|N(\mathit{SB}^\ast)|, \nonumber
\end{eqnarray}
hence the approximation ratio $3\rho H(\Delta +1)+\rho$.
\end{IEEEproof}

\begin{IEEEproof}[\textbf{Proof of Lemma 3}]
Let $S_1 $ be the set $\{\lambda (u,i) \vert u \in d^ + (T_I) \wedge i\in \Upsilon (u,T_I ) \}$. Clearly, $\vert S_1 \vert = \Xi (T_I )$. Furthermore, we have the following statements:
\begin{enumerate}
  \item According to Definition \ref{def:ExtendedGraph}(iii), we know that for any node $u \in d^ + (T_I)$ and any $i_1 ,i_2 \in \Gamma (u), i_1 \ne i_2$, the edge $(\lambda (u,i_1),\lambda (u,i_2 ))$ is in $\widetilde{E}$.
  \item For any two neighboring nodes $v_1 $ and $v_2 $ in $d^ + (T_I)$, there must exit $j_1 \in \Upsilon (v_1 ,T_I ) \cap \Gamma (v_2 )$ and $j_2 \in \Upsilon (v_2,T_I) \cap \Gamma (v_1)$. According to Definition \ref{def:ExtendedGraph}(iv), the edge $(\lambda (v_1 ,j_1 ),\lambda (v_2 ,j_2 ))$ is in $\widetilde{E}$.
  \item For any node $m \in M$, if $m \in d^ + (T_I)$, then for any $i \in \Upsilon (m,T_I )$, $\lambda (m,i)$ is adjacent to $m$, according to Definition \ref{def:ExtendedGraph}(iii). If $m \in d^1(T_I)$, then there must exist a node $m_1 \in d^ + (T_I)$ adjacent to $m$. Since $\Upsilon (m_1 ,T_I)$ is a minimum hitting set of the collection $\{\Gamma (v)\vert v \in nb_{T_I } (m_1 )\}$, there must exist $k_1 \in \Upsilon (m_1 ,T_I ) \cap \Gamma (m)$. Therefore, $\lambda (m_1,k_1 ) \in S_1$. According to Definition \ref{def:ExtendedGraph}(iv), we have $(\lambda (m_1 ,k_1),m) \in \widetilde{E}$.
\end{enumerate}

From 1) and 2) it is easy to know that the sub-graph induced by $S_1 $ in $\widetilde{G}$ is connected. From 3) we know that every node in $M$ is adjacent to a node in $S_1 $. Let $\mathit{SB}_1$ be an arbitrary spanning tree of the sub-graph induced by $S_1$. Then $\mathit{SB}_1$ must be a satellite bridge. Since $\mathit{SB}^\ast$ is a minimum satellite bridge, we have $| N(\mathit{SB}^\ast)| \le |N(\mathit{SB}_1)| = |S_1| = \Xi (T_I )$.
\end{IEEEproof}

\begin{IEEEproof}[\textbf{Proof of Lemma 4}]
Suppose that the nodes in $\mathit{SB}$ belong to $q$ nuclear nodes. Therefore, $N(\mathit{SB})$ can be
partitioned into $q$ mutually disjoint subsets: $A_1 $,$A_2 $,\ldots ,$A_q $,
such that all the nodes in $A_i $ have the same nuclear node $a_i $, and $a_i \ne a_j $ for any $i \ne j, 1 \le i,j
\le q$. Let $A$ be the set $\{a_1 ,a_2 ,...,a_q \}$. According to Definition
\ref{def:MSB}, for any node $u \in M - A$, we can find a node $u' \in N(\mathit{SB})$ adjacent to $u$. So we can connect $u$ to $\mathit{SB}$ by adding the edge $(u,u')$. When all the nodes in $M
- A$ are connected to $\mathit{SB}$, we get a new tree $\mathit{SB}'$.

Then we consider the graph $G' = (V',E')$ whose node set $V'$ is $M \cup A$
and edge set $E'$ is $E'_1 \cup E'_2$, where $E'_1 = \{(a_i ,a_j ) | \exists (u,v) \in \mathit{SB}' \wedge u \in A_i \wedge v \in
A_j \wedge (1 \le i \ne j \le q) \}$ and $E'_2 = \{(a_i ,v) |
\exists (u,v) \in \mathit{SB}' \wedge u \in A_i \wedge v \in (M - A)
\wedge (1 \le i \le q) \}$. From Definition~\ref{def:ExtendedGraph}(iv), it is easy to know that $G'$ is a connected sub-graph of $G$, and all
nodes in $M - A$ are degree-one nodes in $G'$.

Let $R$ be an arbitrary spanning tree of $G'$. Clearly, we have $M \subseteq
N(R)$ and $d^ + (R) \subseteq A$. Let $F(a_i) = \{l | \exists \lambda (a_i
,l) \in A_i \}$ for $1 \le i \le q$. Clearly, $| F(a_i ) | =
| A_i |$. For $a_i \in A$ and $w \in nb_R (a_i)$, we have:
\begin{enumerate}
  \item If $(a_i ,w) \in E'_1$, then there must exist $a_j \in A$ , $u \in A_i$ and $v \in A_j $ such that $w = a_j $ ,$_{ }(u,v) \in \mathit{SB}'$ and $i \ne j$. According to Definition~\ref{def:ExtendedGraph}(iv), there must exist a time slot $l_1 \in \Gamma (a_j)$ such that $u = \lambda (a_i ,l_1 )$. Note that $l_1$ is also in $F(a_i)$. Therefore, $F(a_i) \cap \Gamma (w) = F(a_i) \cap \Gamma (a_j ) \ne \emptyset$.
  \item If $(a_i ,w) \in E'_2$, then there must exist $u \in A_i $ adjacent to $w$. By reasoning similar to 1), we also have $F(a_i ) \cap \Gamma (w) \ne \emptyset $ .
\end{enumerate}
From 1) and 2), we know that any $F(a_i ), 1 \le i \le q$ is a hitting set
of the collection $\{\Gamma (v)  | v \in \mathit{nb}_R (a_i )\}$. Since $d^ + (R) \subseteq A$, we have $\sum_{u \in d^ + (R)} |F(u)| \le \sum_{u \in A}|F(u)| = \sum_{1 \le i \le q} |F(a_i )| = \sum_{1 \le i \le q} |A_i| = |N(\mathit{SB})|$.
\end{IEEEproof}

\begin{IEEEproof}[\textbf{Proof of Theorem 3}]
From Lemma 4 we know that we can find a tree $R^{\ast} $ in $G$ spanning the nodes in $M$ and a function $ F^{\ast} $ such that:
\begin{enumerate}
  \item For any $u \in d^+ (R^\ast)$, $F^\ast(u)$ is a hitting set of the
collection $\{\Gamma (v) | v \in \mathit{nb}_{R^\ast} (u)\}$;
  \item $\sum_{u \in d^+ (R^\ast)} |F^\ast (u)| \le |N(\mathit{SB}^\ast)|$.
\end{enumerate}
Clearly, we also have $\sum_{u \in d^+ (R^\ast)} |\Upsilon (u,R^\ast)| \le \sum_{u \in d^+ (R^\ast)} |F^\ast(u)|$, and
$\Xi (T_I) \le \sum_{u \in d^+ (R^\ast)} |\Upsilon (u,R^\ast)|$. Therefore,
$\Xi (T_I ) \le \sum_{u \in d^+ (R^\ast)} |F^\ast(u)| \le |N(\mathit{SB}^\ast)|$. From Lemma 3 we know $|N(\mathit{SB}^\ast)| \le \Xi (T_I)$. Therefore, $\Xi (T_I) = |N(\mathit{SB}^\ast)|$.
\end{IEEEproof}

\begin{IEEEproof}[\textbf{Proof of Lemma 5}]
From Definition~\ref{def:MEMTCS} we know that $B_{\mathrm{opt}} (u)$ must be a minimum hitting set of the collection $\{\Gamma (v) | v \in \mathit{child}(u,T_{\mathrm{opt}} )\}$. If $u$ is the root of $T_{\mathrm{opt}}$, $\mathit{child}(u,T_{\mathrm{opt}} ) = \mathit{nb}_{T_{\mathrm{opt}}} (u)$. Therefore, we have
$|\Upsilon (u,T_{\mathrm{opt}})| = |B_{\mathrm{opt}}(u)| \le |B_{\mathrm{opt}}(u)| + 1$.

Otherwise, let $u'$ be the parent node of $u$ in $T_{\mathrm{opt}}$. Find an arbitrary $j \in \Gamma (u')$, then $B_{\mathrm{opt}}(u) \cup \{j\}$ must be a hitting set of the collection $\{\Gamma (v) | v \in \mathit{nb}_{T_{\mathrm{opt}}} (u)\}$. So we still have $|\Upsilon (u,T_{\mathrm{opt}} )| \le |B_{\mathrm{opt}} (u) \cup \{j\}| \le
| B_{\mathrm{opt}} (u)| + 1$.
\end{IEEEproof}

\begin{IEEEproof}[\textbf{Proof of Lemma 6}]
If $s \in d^1(T_{\mathrm{opt}})$, then $|B_{\mathrm{opt}} (s)| = 1$. With Lemma 5, we have:
\begin{eqnarray}
\sum_{u \in \mathit{nl}(T_{\mathrm{opt}})} |B_{\mathrm{opt}}(u)| &=& |B_{\mathrm{opt}}(s)| + \sum_{u \in d^+ (T_{\mathrm{opt}} )} |B_{\mathrm{opt}} (u)| \nonumber\\
&\ge& 1 + \sum_{u \in d^+ (T_{\mathrm{opt}})} (\Upsilon (u,T_{\mathrm{opt}} ) - 1) \nonumber \\
&=& \Xi (T_{\mathrm{opt}}) - |d^ +(T_{\mathrm{opt}})| + 1 \nonumber
\end{eqnarray}

If $s \in d^ + (T_{\mathrm{opt}} )$, then $|B_{\mathrm{opt}}(s)| = |\Upsilon(s,T_{\mathrm{opt}} )|$. Therefore, we still have:
\begin{eqnarray}
\!\!\!\!\sum_{u \in \mathit{nl}(T_{\mathrm{opt}})} \!\!\!\!|B_{\mathrm{opt}}(u)| &\!\!=\!\!& |B_{\mathrm{opt}}(s)| + \sum_{u \in d^+ (T_{\mathrm{opt}} )\backslash \{s\}} |B_{\mathrm{opt}} (u)| \nonumber \\
&\!\!\ge\!\!& |\Upsilon (s,T_{\mathrm{opt}})| + \!\!\!\!\!\!\sum_{u \in d^+ (T_{\mathrm{opt}})\backslash \{s\}} \!\!\!\!\!\!(|\Upsilon (u,T_{\mathrm{opt}} )| - 1) \nonumber \\
&\!\!=\!\!& \!\!\!\!\sum_{u \in d^+ (T_{\mathrm{opt}})}\!\! |\Upsilon (u,T_{\mathrm{opt}} )| - (|d^+ (T_{\mathrm{opt}} )| - 1) \nonumber \\
&\!\!=\!\!& \Xi (T_{\mathrm{opt}}) - |d^+ (T_{\mathrm{opt}} )| + 1 \nonumber
\end{eqnarray}

Now $\sum_{u \in \mathit{nl}(T_{\mathrm{opt}} )} |B_{\mathrm{opt}}(u)| \ge \Xi (T_I ) - |d^ + (T_{\mathrm{opt}} )| + 1$ follows from the fact that $\Xi (T_{\mathrm{opt}} ) \ge \Xi (T_I )$.
\end{IEEEproof}

\begin{IEEEproof}[\textbf{Proof of Theorem 4}]
Using Lemma~\ref{lma:optp2}, we have:
\begin{eqnarray}
\!\!&&\!\! \Pi (T_{\mathrm{opt}},B_{\mathrm{opt}}) \nonumber \\
\!\!&=&\!\! \sum_{u \in \mathit{nl}(T_{\mathrm{opt}})} | B_{\mathrm{opt}}(u) | \cdot e_s + \left(| N(T_{\mathrm{opt}})| - 1\right) \cdot e_r \nonumber \\
\!\!&\ge&\!\! (\Xi (T_I) - |d^+ (T_{\mathrm{opt}} )\vert + 1)\cdot e_s +  \left(| N(T_{\mathrm{opt}} )| - 1\right) \cdot e_r \nonumber \\
\!\!&=&\!\! (\Xi (T_I) + 1) \cdot e_s +  \left(| N(T_{\mathrm{opt}} )| - 1\right) \cdot e_r -  |d^ + (T_{\mathrm{opt}} )| \cdot e_s \nonumber
\end{eqnarray}

Since each node in $d^+ (T_{\mathrm{opt}})$ must transmit at least once, we have $|d^+(T_{\mathrm{opt}} )| \cdot e_s \le \Pi(T_{\mathrm{opt}} ,B_{\mathrm{opt}} )$, and hence:
$(\Xi (T_I) + 1) \cdot e_s + (|N(T_{\mathrm{opt}} )| - 1) \cdot e_r \le 2\Pi(T_{\mathrm{opt}} ,B_{\mathrm{opt}} )$.

Let $\alpha = 3\rho H(\Delta + 1) + \rho$. With Lemma~\ref{lma:lbd}, Theorem~\ref{thm:MSB} and Theorem~\ref{thm:eq}, we have:
\begin{eqnarray}
\sum_{u \in d^+ (\overline{T})} |\overline{B}(u)| &=& \sum_{u \in d^+ (\widehat{T})} |\widehat{F}(u)| ~~\le~~ | N(\overline{\mathit{SB}}) | \nonumber \\
&\le& \alpha | N(\mathit{SB}^\ast)| ~~=~~ \alpha \Xi (T_I) \nonumber
\end{eqnarray}

Besides, we have some evident inequalities: $e_r\le e_s$, $|d^1(\overline{T})|\le |M| \le | N(T_{\mathrm{opt}})|$, and $|\overline{B}(u)| \ge 1$ (for any $u \in
d^+(\overline{T}))$. Using these inequalities, we can get:
\begin{eqnarray}
\!\!&&\!\! \Pi (\overline{T},\overline{B}) \nonumber \\
\!\!&=&\!\! \sum_{u \in \mathit{nl}(\overline{T})} |\overline{B}(u)| \cdot e_s +  (| N(\overline{T})| - 1) \cdot e_r \nonumber \\
\!\!&=&\!\! \sum_{u \in d^ + (\overline{T}) \cup \{s\}} |\overline{B}(u)| \cdot e_s + (|d^+ (\overline{T})| + |d^1(\overline{T})| - 1) \cdot e_r \nonumber \\
\!\!&\le&\!\! \left(\sum_{u \in d^+ (\overline{T})} |\overline{B}(u)| + 1\right) \cdot e_s + \!\!\!\!\sum_{u \in d^+ (\overline{T})} e_s + (|M| - 1) \cdot e_r \nonumber \\
\!\!&\le&\!\! \left(2\sum_{u \in d^+ (\overline{T})} |\overline{B}(u)| + 1\right) \cdot e_s + (| N(T_{\mathrm{opt}})| - 1) \cdot e_r \nonumber\\
\!\!&\le&\!\! (2\alpha \cdot \Xi (T_I) + 1) \cdot e_s + (|N(T_{\mathrm{opt}})| - 1) \cdot e_r \nonumber\\
\!\!&\le&\!\! 2\alpha \cdot \left((\Xi (T_I) + 1) \cdot e_s + (|N(T_{\mathrm{opt}})| - 1) \cdot e_r\right) \nonumber\\
\!\!&\le&\!\! 4\alpha \cdot \Pi (T_{\mathrm{opt}} ,B_{\mathrm{opt}} ) \nonumber\\
\!\!&=&\!\! \left(12\rho H(\Delta + 1) + 4\rho\right) \Pi (T_{\mathrm{opt}} ,B_{\mathrm{opt}}), \nonumber
\end{eqnarray}
hence the approximation ratio $12\rho H(\Delta + 1) + 4\rho$.
\end{IEEEproof}



\bibliographystyle{IEEEtran}
\bibliography{IEEEabrv,MyLib}

\begin{thebibliography}{10}
\providecommand{\url}[1]{#1}
\csname url@samestyle\endcsname
\providecommand{\newblock}{\relax}
\providecommand{\bibinfo}[2]{#2}
\providecommand{\BIBentrySTDinterwordspacing}{\spaceskip=0pt\relax}
\providecommand{\BIBentryALTinterwordstretchfactor}{4}
\providecommand{\BIBentryALTinterwordspacing}{\spaceskip=\fontdimen2\font plus
\BIBentryALTinterwordstretchfactor\fontdimen3\font minus
  \fontdimen4\font\relax}
\providecommand{\BIBforeignlanguage}[2]{{%
\expandafter\ifx\csname l@#1\endcsname\relax
\typeout{** WARNING: IEEEtran.bst: No hyphenation pattern has been}%
\typeout{** loaded for the language `#1'. Using the pattern for}%
\typeout{** the default language instead.}%
\else
\language=\csname l@#1\endcsname
\fi
#2}}
\providecommand{\BIBdecl}{\relax}
\BIBdecl

\bibitem{Gu2007}
Y.~Gu and T.~He, ``Data forwarding in extremely low duty-cycle sensor networks
  with unreliable communication links,'' in \emph{Proc. ACM SenSys}, 2007, pp.
  321--334.

\bibitem{Anastasi2009}
G.~Anastasi, M.~Conti, M.~D. Francesco, and A.~Passarella, ``Energy
  conservation in wireless sensor networks: A survey,'' \emph{Ad Hoc Networks},
  vol.~7, no.~3, pp. 537--568, 2009.

\bibitem{Wang2009}
F.~Wang and J.~Liu, ``Duty-cycle-aware broadcast in wireless sensor networks,''
  in \emph{Proc. IEEE INFOCOM}, 2009, pp. 468--476.

\bibitem{Guo2009}
S.~Guo, Y.~Gu, B.~Jiang, and T.~He, ``Opportunistic flooding in low-duty-cycle
  wireless sensor networks with unreliable links,'' in \emph{Proc. ACM
  MobiCom}, 2009, pp. 133--144.

\bibitem{Su2009}
L.~Su, B.~Ding, Y.~Yang, T.~F. Abdelzaher, G.~Cao, and J.~C. Hou, ``ocast:
  Optimal multicast routing protocol for wireless sensor networks,'' in
  \emph{Proc. IEEE ICNP}, 2009, pp. 151--160.

\bibitem{Hong2010}
J.~Hong, J.~Cao, W.~Li, S.~Lu, and D.~Chen, ``Minimum-transmission broadcast in
  uncoordinated duty-cycled wireless ad hoc networks,'' \emph{{IEEE} Trans.
  Veh. Technol.}, vol.~59, no.~1, pp. 307--318, 2010.

\bibitem{Xiong2011}
S.~Xiong, J.~Li, M.~Li, J.~Wang, and Y.~Liu, ``Multiple task scheduling for
  low-duty-cycled wireless sensor networks,'' in \emph{Proc. IEEE INFOCOM},
  2011, pp. 1323--1331.

\bibitem{Gui2004}
C.~Gui and P.~Mohapatra, ``Power conservation and quality of surveillance in
  target tracking sensor networks,'' in \emph{Proc. ACM MobiCom}, 2004, pp.
  129--143.

\bibitem{He2006}
T.~He, P.~Vicaire, T.~Yan, Q.~Cao, G.~Zhou, L.~Gu, L.~Luo, R.~Stoleru, J.~A.
  Stankovic, and T.~F. Abdelzaher, ``Achieving long-term surveillance in
  vigilnet,'' in \emph{Proc. IEEE INFOCOM}, 2006, pp. 1--12.

\bibitem{Mo2009}
L.~Mo, Y.~He, Y.~Liu, J.~Zhao, S.~Tang, X.-Y. Li, and G.~Dai, ``Canopy closure
  estimates with greenorbs: sustainable sensing in the forest,'' in \emph{Proc.
  ACM SenSys}, 2009, pp. 99--112.

\bibitem{Xing2009}
G.~Xing, M.~Li, H.~Luo, and X.~Jia, ``Dynamic multiresolution data
  dissemination in wireless sensor networks,'' \emph{{IEEE} Trans. Mobile
  Comput.}, vol.~8, pp. 1205--1220, 2009.

\bibitem{Wieselthier2000}
J.~Wieselthier, G.~Nguyen, and A.~Ephremides, ``On the construction of
  energy-efficient broadcast and multicast trees in wireless networks,'' in
  \emph{Proc. IEEE INFOCOM}, 2000, pp. 585--594.

\bibitem{Wan2004}
P.-J. Wan, G.~Calinescu, and C.-W. Yi, ``Minimum-power multicast routing in
  static ad hoc wireless networks,'' \emph{{IEEE/ACM} Trans. Netw.}, vol.~12,
  no.~3, pp. 507--514, 2004.

\bibitem{Liang2006}
W.~Liang, ``Approximate minimum-energy multicasting in wireless ad hoc
  networks,'' \emph{{IEEE} Trans. Mobile Comput.}, vol.~5, no.~4, pp. 377--387,
  2006.

\bibitem{Li2007}
D.~Li, Q.~Liu, X.~Hu, and X.~Jia, ``Energy efficient multicast routing in ad
  hoc wireless networks,'' \emph{Computer Communications}, vol.~30, no.~18, pp.
  3746--3756, 2007.

\bibitem{Liang2009}
W.~Liang, R.~Brent, Y.~Xu, and Q.~Wang, ``Minimum-energy all-to-all
  multicasting in wireless ad hoc networks,'' \emph{{IEEE} Trans. Wireless
  Commun.}, vol.~8, no.~11, pp. 5490--5499, 2009.

\bibitem{Charikar1998}
M.~Charikar, C.~Chekuri, T.-y. Cheung, Z.~Dai, A.~Goel, S.~Guha, and M.~Li,
  ``Approximation algorithms for directed steiner problems,'' in \emph{Proc.
  ACM-SIAM SODA}, 1998, pp. 192--200.

\bibitem{Garey1979}
M.~R. Garey and D.~S. Johnson, \emph{Computers and Intractability: A Guide to
  the Theory of NP-Completeness}.\hskip 1em plus 0.5em minus 0.4em\relax New
  York: W. H. Freeman, 1979.

\bibitem{Ausiello1980}
G.~Ausiello, A.~D'Atri, and M.~Protasi, ``Structure preserving reductions among
  convex optimization problems,'' \emph{Journal of Computer and System
  Sciences}, vol.~21, no.~1, pp. 136--153, 1980.

\bibitem{Feige1998}
U.~Feige, ``A threshold of ln\textit{n} for approximating set cover,'' \emph{J.
  ACM}, vol.~45, pp. 634--652, July 1998.

\bibitem{Cormen2001}
T.~H. Cormen, C.~E. Leiserson, R.~L. Rivest, and C.~Stein, \emph{Introduction
  to Algorithms, Second Edition}.\hskip 1em plus 0.5em minus 0.4em\relax MIT
  Press and McGraw-Hill, 2001.

\bibitem{Mehlhorn1988}
K.~Mehlhorn, ``A faster approximation algorithm for the steiner problem in
  graphs,'' \emph{Information Processing Letters}, vol.~27, no.~3, pp.
  125--128, 1988.

\bibitem{Liang2000}
B.~Liang and Z.~J. Haas, ``Virtual backbone generation and maintenance in ad
  hoc network mobility management,'' in \emph{Proc. IEEE INFOCOM}, 2000, pp.
  1293--1302.

\bibitem{Bauer1996}
F.~Bauer and A.~Varma, ``Distributed algorithms for multicast path setup in
  data networks,'' \emph{{IEEE/ACM} Trans. Netw.}, vol.~4, no.~2, pp. 181--191,
  1996.

\bibitem{Chalermsook2005}
P.~Chalermsook and J.~Fakcharoenphol, ``Simple distributed algorithms for
  approximating minimum steiner trees,'' in \emph{Proc. COCOON}, 2005, pp.
  380--389.

\bibitem{Muhammad2006}
R.~B. Muhammad, ``Distributed steiner tree algorithm and its application in
  ad-hoc wireless networks,'' in \emph{Proc. ICWN}, 2006, pp. 173--178.

\bibitem{Makki1996}
S.~A.~M. Makki and G.~Havas, ``Distributed algorithms for depth-first search,''
  \emph{Information Processing Letters}, vol.~60, no.~1, pp. 7--12, 1996.

\bibitem{Ruiz2005}
P.~M. Ruiz and A.~F. Gomez-Skarmeta, ``Approximating optimal multicast trees in
  wireless multihop networks,'' in \emph{Proc. ISCC}, 2005, pp. 686--691.

\bibitem{Ruiz2007}
P.~M. Ruiz and I.~Stojmenovic, ``Cost-efficient multicast routing in ad hoc and
  sensor networks,'' in \emph{Handbook on Approximation Algorithms and
  Metaheuristics}, T.~F. Gonzalez, Ed.\hskip 1em plus 0.5em minus 0.4em\relax
  Chapman \& Hall/CRC, 2007, pp. 65--1.

\bibitem{Kou1981}
L.~Kou, G.~Markowsky, and L.~Berman, ``A fast algorithm for steiner trees,''
  \emph{Acta Informatica}, vol.~15, pp. 141--145, 1981.

\bibitem{Johnson1973}
D.~S. Johnson, ``Approximation algorithms for combinatorial problems,'' in
  \emph{Proc. ACM STOC}, 1973, pp. 38--49.

\end{thebibliography}
%
%
%

\end{document}